# Work Function Trends and New Low Work Function Boride and Nitride Materials for Electron Emission Applications

Tianyu Ma[1], Ryan Jacobs[1], John Booske[2], Dane Morgan[1*]

[1]Department of Materials Science and Engineering, University of Wisconsin-Madison, Madison, WI, USA.

[2]Department of Electrical and Computer Engineering, University of Wisconsin-Madison, Madison, WI, USA.

*To whom correspondence should be addressed. Email: ddmorgan@wisc.edu

**Abstract**

LaB$_6$ has been used as a commercial electron emitter for decades. Despite the large number of studies on the work function of LaB$_6$, there is no comprehensive understanding of work function trends in the hexaboride materials family. In this study, we use Density Functional Theory (DFT) calculations to calculated trends of rare earth hexaboride work function and rationalize these trends based on the electronegativity of the metal element. We predict that alloying LaB$_6$ with Ba can further lower the work function by ~0.2 eV. Interestingly, we find that alloyed (La, Ba)B$_6$ can have lower work functions than either LaB$_6$ or BaB$_6$, benefitting from an enhanced surface dipole due to metal element size mismatch. In addition to hexaborides we also investigate work function trends of similar materials families, namely tetraborides and transition metal nitrides, which, like hexaborides, are electrically conductive and refractory and thus may also be promising materials for electron emission applications. We find that tetraborides consistently have higher work functions than their hexaboride analogues as the tetraborides having less ionic bonding and smaller positive surface dipoles. Finally, we find that HfN has a low work function of about 2.2 eV, making HfN a potentially promising new electron emitter material.

## 1. Introduction

The ability to easily emit electrons from a material into vacuum forms the linchpin of numerous modern technologies ranging from medical imaging equipment to satellite communications.[1–5] Materials used as thermionic electron emitters broadly fall into two groups: dispenser cathodes and monolithic cathode materials.[6] For all of these materials, a main figure of merit is the work function of the material, i.e. the energy required to liberate an electron bound in the material and move that electron into vacuum, where a lower work function value is typically desired.[7,8] Lower work functions are desired because many electron emission applications require high emitted current densities or high brightness from the electron source. Thermionic emitted current densities depend exponentially on the work function by virtue of the Richardson-Laue-Dushman (RLD) equation, so even small (e.g. 0.1 eV scale) reductions in the work function may result in desirable improvements to device performance.[9] Dispenser cathodes, such as the classic Ba-impregnated cathode and newer scandate cathode, typically consist of a porous tungsten body infiltrated with an alkaline-earth containing emission mix, e.g. BaO combined with other metal oxides.[6] At high temperature, the emission mix decomposes, and mobile Ba ions migrate to the surface, where they bind with oxygen and form work-function-lowering dipole species.[10–12] Monolithic materials, by contrast, consist of a single, non-porous material, which natively contains low work function surfaces without the presence of additional work function-lowering species. For many applications, the chief advantages of employing a monolithic material are improved lifetime, more reliable device performance (the volatile surface species can desorb and cause performance issues in many devices), and more straightforward material processing and activation.[13] The main disadvantage of using monolithic materials is that they tend to have higher work functions than dispenser cathodes (e.g. about 0.5-1 eV higher, depending on the precise materials considered).[6]

A seminal event in the development of modern monolithic thermionic emitters came in 1951, when Lafferty, studying the electron emission properties of several hexaborides, found that lanthanum hexaboride ($LaB_6$) displayed outstanding thermionic emission properties, with a low measured work function of about 2.6 eV.[14] This discovery by Lafferty later spurred tremendous interest in this material space, with many following studies making progress in $LaB_6$ materials processing to successfully synthesize single crystal $LaB_6$, in-depth studies of surface-dependent work function to further assess emission performance and guide single crystal engineering,[15,16] and stability studies which found that $LaB_6$ has low vapor pressure at high temperature, ensuring sufficient material stability at elevated temperatures for thousands of hours of collective emission.[16,17] As a result of these attractive materials properties, $LaB_6$ has been used for decades as a commercial electron source in applications needing a combination of high brightness and long life, such as electron microscopes, lithography systems, electron-beam welders, X-ray sources, and free electron lasers.[18,19]

Despite the success of $LaB_6$ as a commercial material, new monolithic electron emitters with lower work function are still desirable to further improve the performance of devices using electron emitters. Besides a low work function, monolithic materials that function as thermionic emitters need to have a high melting temperature (to facilitate high temperature stability) and high electrical conductivity (to facilitate adequate supply of electrons for emission). With these key properties in mind, there are additional interesting materials families beyond hexaborides that we explore in this work. These additional materials families consist of rare-earth tetraborides[20,21] and transition metal nitrides (TMNs). We note here that transition metal carbides (TMCs) may also be interesting to explore, but previous computational studies on TMCs[22,23] show work functions generally higher than 3 eV. Therefore, we do not perform further comprehensive studies of TMC work functions in this study.

Regarding rare earth tetraborides, some experimental studies on the thermionic emission properties of polycrystalline $YB_4$ and $GdB_4$ showed low work functions of 2.38 eV and 1.45 eV, respectively. However, these studies obtained the work function by fitting both the work function and Richardson constant in the RLD equation, where the resulting Richardson constants were quite small (0.1 and $10^{-3}$, compared to the theoretical value of 120 $A \cdot cm^{-2} \cdot K^{-2}$) and only small current densities were observed, indicating they are likely inferior to $LaB_6$.[24] Despite these results indicating that $YB_4$ and $GdB_4$ are lower quality emitters than $LaB_6$, it is still possible that there exists a promising new cathode material among other tetraboride compositions. Regarding TMNs, these materials are also conductive and refractory,[25,26] therefore, we further investigated this class of materials to assess if any TMN materials have low work functions and be useful in electron emission applications. While previous computational studies showed that TMCs have work functions around 3 eV, to our knowledge there is no systematic study on TMNs. Qualitatively, we might expect TMNs to have lower work functions than TMCs because the Fermi level is elevated by doping of an extra electron from N compared to C. While some experimental measurements on TMN work functions did not yield low work functions,[27,28] such experiments were carried out on TMNs with mixed (001) and (111) surfaces. Electron emitters with highly heterogeneous or patchy emitting surfaces may result in a measured work function higher than the desired low work function surface, so isolation of single (001) or (111) terminated TMNs may yield low work functions if either of these surfaces possesses a low work function.

As one of the main electron emitters used in numerous commercial applications, $LaB_6$ has remained essentially unchanged for decades. Despite the large number of studies on hexaboride emitters, they are often limited to one or a few hexaborides, and different studies adopt different sample preparation conditions and/or measurement methods. These differences can make one-to-one comparison of results difficult, and make chemical trends or effects of different surfaces challenging to discern. Importantly, to our knowledge there is no study which comprehensively evaluates and provides understanding of work function trends of the set of hexaboride, tetraboride, and TMN materials that are likely to form. Such a study would reveal trends of material composition with work function and enable understanding of coupling between composition, structure and work function. Knowledge of materials trends and understanding of the relationship between composition, structure and work function may suggest possible avenues toward designing new conductive, refractory, boride and nitride materials with lower work function than $LaB_6$.

In this work, we have conducted Density Functional Theory (DFT) simulations on the full rare earth series of pure hexaborides, along with select alkaline earth hexaborides and alloyed hexaborides. In addition, we have performed DFT simulations on all rare-earth tetraborides. Although Eu does not form $EuB_4$ due to its preference for divalent configuration[21], we still included it in our simulations for completeness and to help observe trends with composition. Finally, we have performed DFT simulations on a series of (001)- and (111)-surfaces of TMNs. From our results, $LaB_6$ has a work function of 2.01 eV, which is the lowest among all pure rare-earth hexaborides, and we find that the $La_{0.25}Ba_{0.75}B_6$ alloy has an even lower work function of 1.84 eV. In addition, we find the work function of tetraborides has a similar trend as that of hexaborides, but the work function of each tetraboride is higher than its hexaboride analogue, and thus tetraborides are likely less interesting for low work function electron emission applications. For TMNs, we find the relationship between the work functions of different surfaces or different transition metals is generally consistent with that in TMCs from previous studies. We also find that the (001) surface of HfN has a low work function of 2.16 eV and thus may also be an interesting material to further explore for possible emission applications.

The chemical trends in the work function across all the systems studied are in agreement with chemical intuition based on electronegativity. Specifically, more electropositive metals have electrons that can be removed more easily, and therefore result in lower work function materials. We find that the more electropositive elements on the left side of the periodic table tend to result in lower work function materials, and the work function tends to increase proceeding to the right on the periodic table. These trends persist for hexaborides, tetraborides, and the TMN materials examined here. Work function changes resulting from alloying the hexaborides are understood in terms of electronegativity, size, and surface dipole effects.

## 2. Computational methods

All calculations were performed with DFT using the *Vienna Ab Initio Simulation Package* (VASP)[29] software using the Generalized Gradient Approximation of Perdew-Burke-Ernzerhof (GGA-PBE)[30] functional. The cut-off energy for the plane-wave basis set was 500 eV. The convergence criterion of ionic steps is $10^{-3}$ eV per supercell energy change. All calculations were performed with spin polarization enabled. PBE-type pseudopotentials utilizing the projector augmented wave (PAW)[31] method were used for all atoms. For bulk calculations, hexaborides (space group $Pm\bar{3}m$) used a cell consisting of one formula unit containing 7 atoms for pure compounds, and an enlarged supercell consisting of eight unit cells containing 56 atoms ($2a_0 \times 2a_0 \times 2a_0$) for alloyed compounds to accommodate alloy components, with a Monkhorst-Pack[32] $12 \times 12 \times 12$ $k$-point mesh and $6 \times 6 \times 6$ $k$-point mesh, respectively. Tetraborides (space group P4/mbm) used a cell consisting of four formula units containing 20 atoms, with a Monkhorst-Pack $7 \times 7 \times 12$ $k$-point mesh. TMNs (space group $Fm\bar{3}m$) used a supercell consisting of four formula units containing 8 atoms, with a Monkhorst-Pack $12 \times 12 \times 12$ $k$-point mesh.

To obtain the work function, we performed symmetric slab calculations with a vacuum region at least 20 Å thick, and $k$-point meshes which were scaled from the bulk values above according to the dimensions of each slab supercell. Hexaboride slabs are $5a_0$ in thickness, tetraboride slabs are $4a_0$ in thickness. We study the (001) metal-terminated surface, with the top two layers of metal atoms and one layer of the boron cage allowed to relax. The (001) surface can be terminated by metal or B. The (001)-metal terminated surface is generally believed to be the lowest work function surface in hexaborides based on experimental observations of (001) surfaces and simple dipole arguments which imply that a metal vs. B termination will have lower work function.[33–37] While there is little information in the literature on observed tetraboride terminations, we assumed the (001)-metal terminated surface is the most relevant and lowest work function based on the similarity between hexaborides and tetraborides. TMN slabs are $4a_0$ in thickness and top three layers are relaxed. We examined (001) and (111) surfaces of TMNs as they have been experimentally observed.[27,28] **Figure 1** shows pictures of the bulk and surface slab structures for a representative composition of each class of material examined in this work.

We note that all calculations are performed on perfect stoichiometric systems in that no vacancies are included. This is potentially a significant approximation for some TMNs, which have been observed to have N vacancies at high concentration (>10%) under some conditions.[38–40]

However, stoichiometric TMNs are accessible in experiments for some synthesis methods,[41] so our results would be directly applicable to those materials, and small vacancy concentrations are expected to have a small effect on work function for metallic systems. For cases with large concentrations of N vacancies the situation can be quite complex. N vacancies can have a wide range of concentrations and long range order and/or surface segregation.[42] In addition, the influence of N vacancies may not be monotonic.[39,43] All of these factors make properly including vacancy effects into the work function calculations potentially very complicated and beyond the scope of the present study, although an interesting topic for future work.

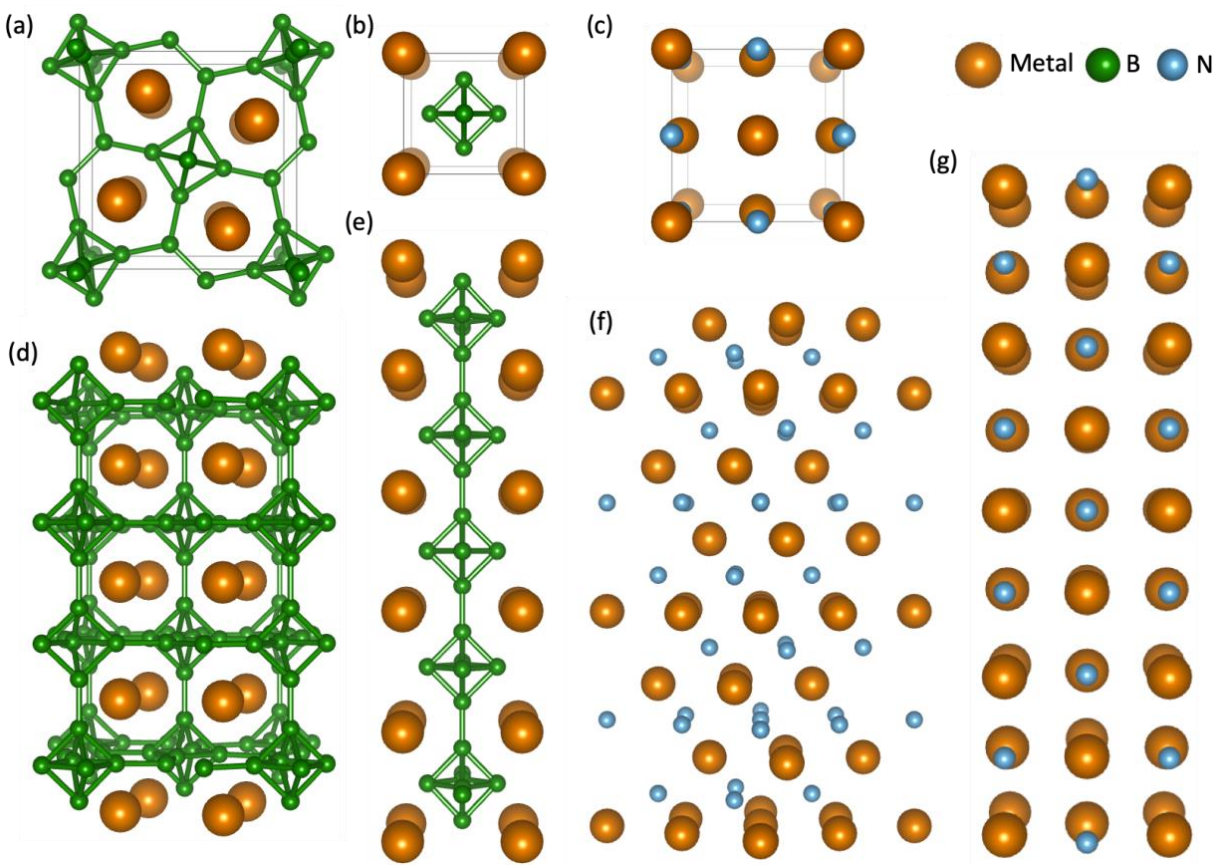

**Figure 1. Supercells of bulk and slab hexaboride, tetraboride, and transition metal nitride used in our calculations. Bulk tetraboride, hexaboride, and TMN are shown in (a), (b), (c), respectively; (d) (001)-metal terminated tetraboride; (e) (001)-metal terminated hexaboride; (f) (111)-metal terminated TMN; (g) (001) terminated TMN.**

## 3. Results and discussion

### 3.1 Review of hexaboride studies

Since its discovery in 1951, many experimental and computational studies have been performed on rare earth hexaborides, especially $LaB_6$, to study their work functions.[15,16,50–59,17,18,44–49] Here, we begin by reviewing the available data from experimental studies from 1951 to present which aimed to measure $LaB_6$ work function, as shown below in **Fig. 2(a)**.

In **Fig. 2(a)** we summarize work function data which originate from different sample morphologies: polycrystalline $LaB_6$ and (001) surface of single crystalline $LaB_6$; and different measurement techniques: thermionic emission (denoted as TE), photoelectron emission (PE), or field emission (FE). In **Fig. 2(b)**, we condense the data from **Fig. 2(a)** to provide averages and ranges of the work function data from different measurement methods. For thermionic emission, both polycrystal and single crystal averages are around 2.6 eV, although polycrystalline work functions seem to be more scattered, with a 0.8 eV range. This observed larger range for polycrystalline work functions is mainly caused by a single work function measurement of 1.98 eV measured by Zhou *et al.*[57], which is notably lower than other measurements. In the study from Zhou *et al.*, the authors fitted both the Richardson constant *A* and the work function in the RLD equation, resulting in a small $A = 1.24 \text{ A} \cdot \text{cm}^{-2} \cdot \text{K}^{-2}$. The authors performed another fit using the theoretical $A = 120 \text{ A} \cdot \text{cm}^{-2} \cdot \text{K}^{-2}$, and the corresponding work function was 2.69 eV, very close to the average reported thermionic work function shown in **Fig 2(b)**. We therefore believe that the very low *A* and work function values from Zhou *et al.* are a result of uncertainty in the fitting and a strong coupling between these two variables which can occur during fitting, leading to both values being quite far off and yet fitting well to physical emission data. We believe that Zhou *et al.*'s work function obtained from their fit with the theoretical *A* value is likely more accurate. In **Fig 2(b)**, the average work function values from multiple studies are in very good agreement between thermionic emission of polycrystals, thermionic emission of single crystals, and field emission measurements, all of which give an average value of about 2.6 eV. The photoelectron emission measured averaged work functionat 2.3 eV is generally lower than the work functions from other measurements, but the range of reported values overlaps with the range of values from both thermionic and field emission measurements, again indicating overall good agreement with other experiments.

In addition to the experimental LaB$_6$ work function data, there are also computational studies on LaB$_6$ predicting work functions of 2.07 eV[60] and 2.27 eV[61]. Given the observed spread in the experimental data, it is difficult to make one-to-one comparisons between calculation and experiment, however the computational results are generally lower than the experimental results. The origin of this discrepancy is not clear but discussed further in Sec. 3.2. LaB$_6$ surfaces other than the (001) surface have also been studied. While one study showed the (011) surface has an even lower work function (2.68 eV) than (001) surface (2.86 eV),[51] there is historically more evidence both from experiments and computation to suggest that the (001) surface has the lowest work function.[15,54,60] There are also studies on other hexaboride compositions, including BaB$_6$, YB$_6$, CeB$_6$, PrB$_6$, NdB$_6$, SmB$_6$, EuB$_6$, and GdB$_6$, and their alloys with rare-earth or transition metals in search of better emitters.[24,47,62–68] However, none of them achieved higher current density than LaB$_6$.

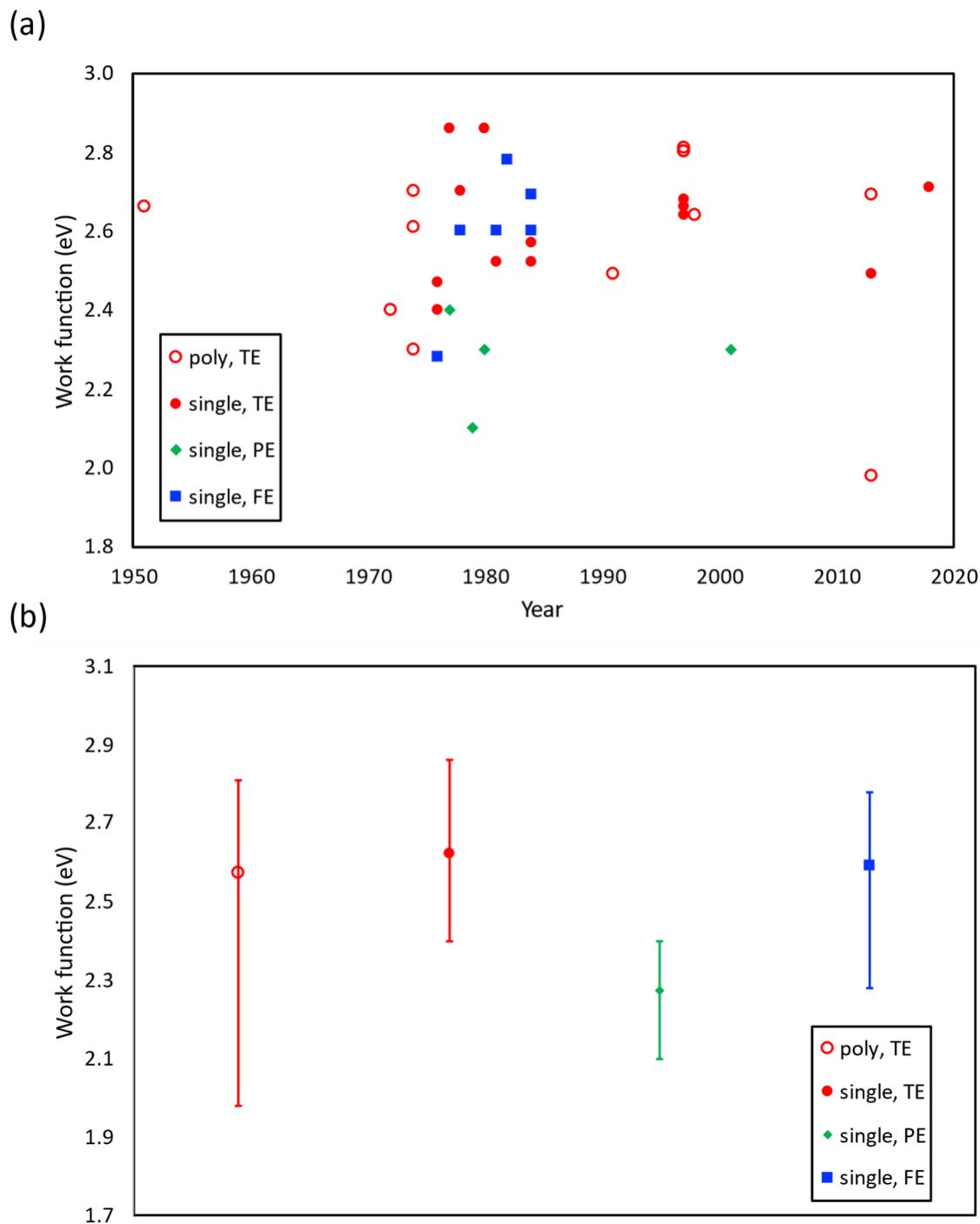

**Figure 2. Experimental work functions of LaB$_6$ from different studies.**[15,16,50–59,17,18,44–49] **TE denotes thermionic emission, PE denotes photoelectron emission, and FE denotes field emission. Measurements on single crystals are all from (001) surface. Error bars denote the ranges of the work function data, derived from taking the difference between the maximum and minimum values measured in experiments.**

## 3.2 Hexaboride work function trends and effects of alloying

The work functions of (001)-metal-terminated rare earth hexaborides from our calculations are shown in **Fig. 3**, along with experimental work function values for each material, where available. For the most studied hexaboride, LaB$_6$, our DFT-calculated work function is generally lower than experimentally measured values, and the work function of LaB$_6$ from our calculation (2.01 eV) is consistent with previous computational studies, which found values of 2.07 eV[60] and 2.27 eV[61]. The observed discrepancy between calculated and experimental LaB$_6$ work function may be caused by factors related to measurement method or method of data analysis (e.g. fitting work function or work function and Richardson constant in thermionic emission experiments), or materials-related factors such as different stoichiometries, defect concentrations, and surface conditions determined by sample preparation. In addition, it is possible that errors in DFT play a role, as previous studies benchmarking the accuracy of DFT calculated work functions suggest DFT-GGA level work functions are ~0.3 eV lower than experimental values for metals,[69,70] and a separate study of DFT-HSE level work functions have a ~0.2 eV deviation on average from experiments for SrTiO$_3$.[71] With this combination of factors, it is difficult here to judge whether DFT or experimental measurement is more accurate, or to fully understand the origin of the observed discrepancies between the calculated and experimental work functions in **Figure 3**. Nevertheless, it is apparent that the trend of work function with composition is qualitatively the same between experiment and DFT. Therefore, we believe that the agreement between experiments and DFT is close enough that DFT guidance on trends is likely to be relevant for the performance of real materials.

In **Fig. 3**, we also plot the average experimental values for each composition for which data was available. We find that the average (standard deviation) shift between our DFT calculated work functions and the experimental values is 1.11 (0.45) eV, mainly due to the large spread in experimental data for less-studied materials, particularly NdB$_6$ and EuB$_6$. If we restrict our comparison between DFT and experiment to the most studied hexaborides (i.e. LaB$_6$, CeB$_6$, PrB$_6$, and SmB$_6$), the deviation between DFT and experiment is 0.88 eV on average, and is 0.65 eV on average if one considers the most recent measurement of SmB$_6$ by Waldhauser *et al.*[47] to be the most reliable. In this latter comparison, the average error between DFT and experiment of 0.65 eV is reasonably uniform across the compositions considered, providing an approximate average error between DFT and experiment for the most-studied hexaboride materials. Again we note that although our calculated work functions appear to be lower than experiments, the accuracy of the trends suggests that a new material may still be a promising emitter if our calculations yield a work function lower than our calculated value for LaB$_6$.

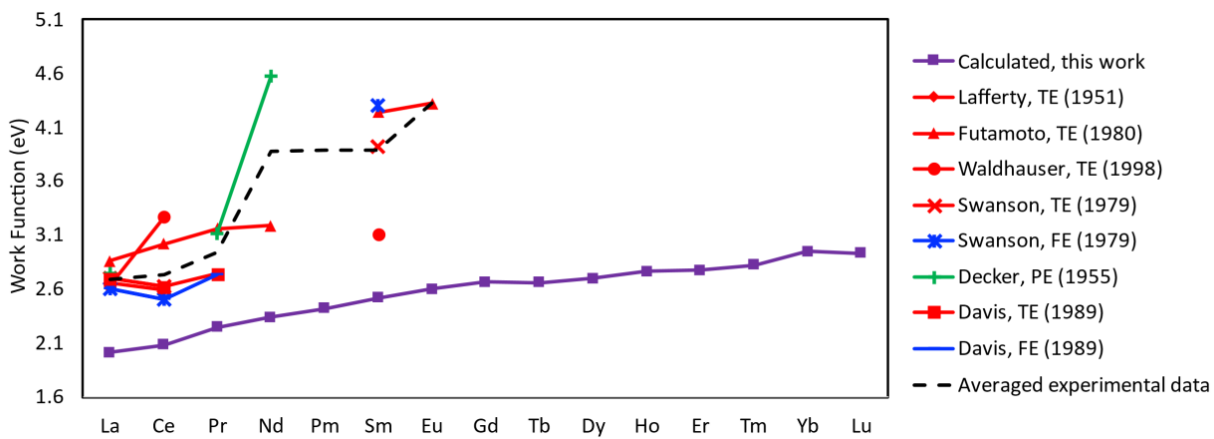

**Fig 3. DFT-calculated work functions of hexaborides, comparing to experimental work functions. Thermionic emission data are in red, photoelectron emission data are in green, field emission data are in blue, and averaged experimental data are denoted as black dashed line.**

From our calculations, LaB$_6$ has the lowest work function among all pure rare-earth hexaborides, and the work function increases as the parent rare-earth metal moves from the left to the right on the periodic table. These results can be rationalized by assuming the work function value is being governed by the electronegativity of the parent metal, which increases when proceeding from left to right on the periodic table. Parent metals that are less electronegative (i.e. more electropositive) will bind electrons less strongly, possibly resulting in both weaker overall binding of electrons in the bulk of the material and a stronger positive surface dipole which acts to produce a low work function (recall the (001) surface of hexaborides is a layer of metal atoms, which will be positively charged). This observed trend also makes sense as a positive correlation between work function and electronegativity has been reported and validated for elements[72] and binary compounds[73]. Therefore, according to our results, it makes sense that LaB$_6$ has the lowest work function among all pure rare-earth hexaborides.

There have also been experimental studies that attempted to alloy LaB$_6$ with other rare-earth or transition metals, but most such experiments of alloyed hexaborides did not yield higher current density than LaB$_6$. From our electronegativity arguments made above, it would seem plausible that hexaboride alloys incorporating some fraction of alkaline earth metal, as opposed to other rare earth or transition metals, may yield a lower work function and thus result in higher emitted current densities than the parent rare earth hexaboride. To test this hypothesis, we chose to alloy Ba into LaB$_6$, as Ba is the most electropositive stable alkaline earth metal, and calculated the work functions of a series of La$_x$Ba$_{1-x}$B$_6$ alloys.

We considered seven different La$_x$Ba$_{1-x}$B$_6$ alloys (x = 0.125, 0.25, 0.375, 0.6, 0.625, 0.75 and 0.875). The structures of the slabs of these compounds are shown in **Fig. 4**, where La$_{0.125}$Ba$_{0.875}$B$_6$ and La$_{0.875}$Ba$_{0.125}$B$_6$ have two possible terminations, both of which are included in our calculations. The calculated work functions of these La$_x$Ba$_{1-x}$B$_6$ alloys are shown in **Fig. 5**. As we hypothesized, most of these alloys have work functions lower than pure LaB$_6$, where La$_{0.25}$Ba$_{0.75}$B$_6$ has the lowest work function of 1.84 eV, which is about 0.2 eV lower than pure LaB$_6$. From Bader charge analysis, we confirm that there is charge transfer between La/Ba and B, and in La$_x$Ba$_{1-x}$B$_6$ the charge transfer between La and B is stronger than that in pure LaB$_6$, resulting in a more electropositive surface layer, as shown below in **Table 1**. Further examination on the relaxed surface structure of La$_x$Ba$_{1-x}$B$_6$ reveals that Ba alloying also creates more surface rumpling with Ba protruding outward due to the misfit between La and Ba atomic radii, as shown in **Fig 4**. Both effects produce a larger surface dipole that facilitates electron emission. If one were relying on the argument of metal electronegativity alone, this would suggest that BaB$_6$ should have the lowest work function. However, we find that pure BaB$_6$ does not have a lower work function than the La$_x$Ba$_{1-x}$B$_6$ alloys. This fact indicates that relying on metal electronegativity alone is not sufficient to understand the work function of hexaborides, where, for alloyed cases, the size mismatch between the alloyed metal atoms also plays an important role.

**Table 1. Charge transfer of surface and subsurface Ba/La atoms in La$_x$Ba$_{1-x}$B$_6$, LaB6 and BaB$_6$.**

| Material | $La_xBa_{1-x}B_6$ | | $LaB_6$ | $BaB_6$ |
|---|---|---|---|---|
| Surface atom | Ba | La | La | Ba |
| Charge transfer (e) | 1.31 | 1.79 | 1.50 | 1.32 |
| Subsurface atom | Ba | La | La | Ba |
| Charge transfer (e) | 1.36 | 1.59 | 1.67 | 1.41 |

In addition to the work function of alloyed hexaborides, we examined a basic measure of alloy stability by calculating the mixing energies of alloying $LaB_6$ and $BaB_6$:

$$x\mathrm{LaB_6} + (1-x)\mathrm{BaB_6} \rightarrow \mathrm{La_xBa_{1-x}B_6}$$

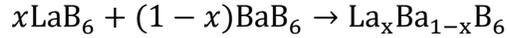

Based on our calculations, the mixing energies of various concentrations of $La_xBa_{1-x}B_6$ are all quite close to zero, and slightly negative for the La range of 12.5-50%, which is the range of lowest alloy work functions. We speculate the mixing energies are near zero or weakly negative because the metal atoms are isolated from each other by the large boron frames, and thus have little interaction between them. If one further considers the effect of the mixing entropy, these results predicts it should be possible to experimentally synthesize such alloys, consistent with the results of Hasan *et al.*[64] who successfully synthesized a series of $La_xBa_{1-x}B_6$ at different concentrations, some of which formed stable alloys. Since the work function depends strongly on surface configuration, we also explored the effect of different La/Ba arrangements on the work function for $La_{0.25}Ba_{0.75}B_6$, which has the lowest work function among all $La_xBa_{1-x}B_6$ alloys we studied. The results of considering all possible La and Ba arrangements within the top three layers of our simulation cell are shown in **Fig 5(b)**. The range of values is about 0.1 eV and the general trend indicates that lower work functions also correspond to more stable surface configurations. In **Fig 5(b)**, the configuration that corresponds to $La_{0.25}Ba_{0.75}B_6$ work function data point in **Fig 5(a)** is colored in blue. This configuration has neither the lowest work function nor the lowest surface energy, although it is close to both lower limits. This surface configuration is chosen to show in **Fig 5(a)** to be consistent with other $La_xBa_{1-x}B_6$ structures used in **Fig 5(a)** in which the structure is generated by placing the alloying atoms as far from each other as possible.

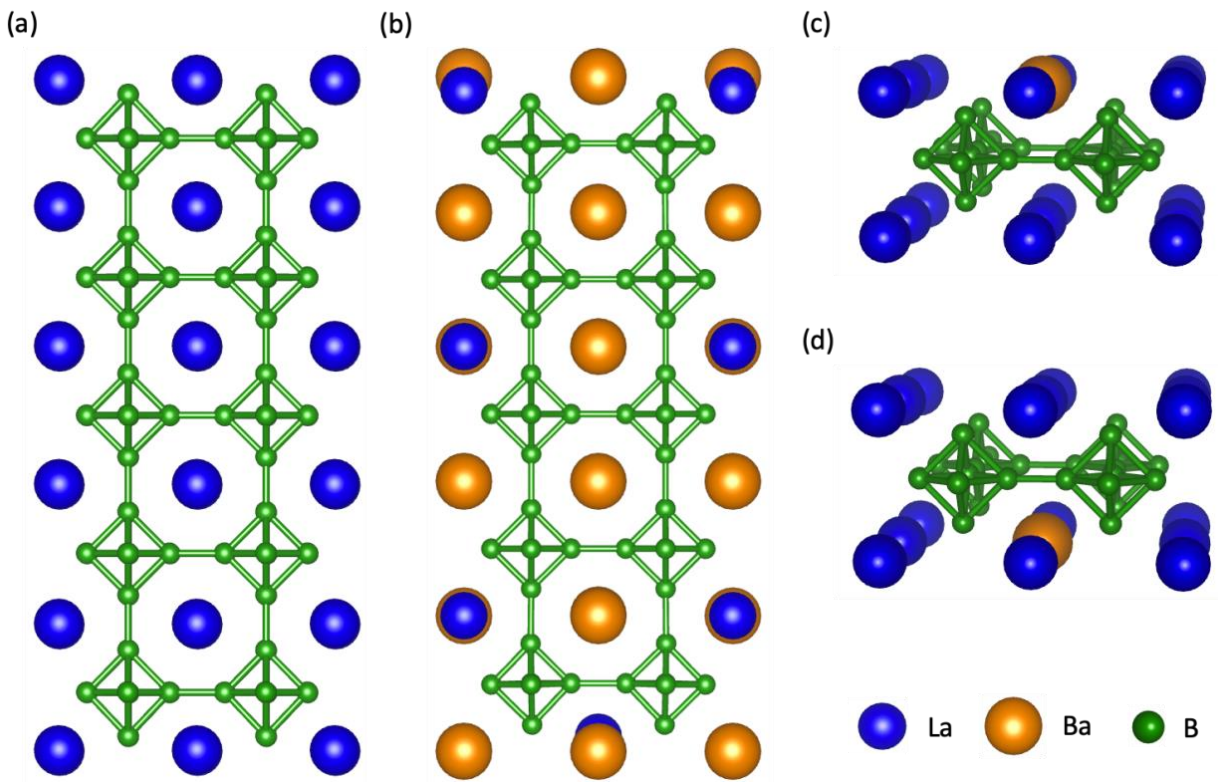

**Fig 4. Relaxed structures of LaB$_6$ and La$_{0.25}$Ba$_{0.75}$B$_6$ slabs, and La$_{0.875}$Ba$_{0.125}$B$_6$ surfaces showing two possible configurations. (a) relaxed LaB$_6$ slab; (b) relaxed La$_{0.25}$Ba$_{0.75}$B$_6$ slab exhibiting surface rumpling; top three layers of La$_{0.875}$Ba$_{0.125}$B$_6$ slabs with (c) Ba rich termination, and (d) La rich termination.**

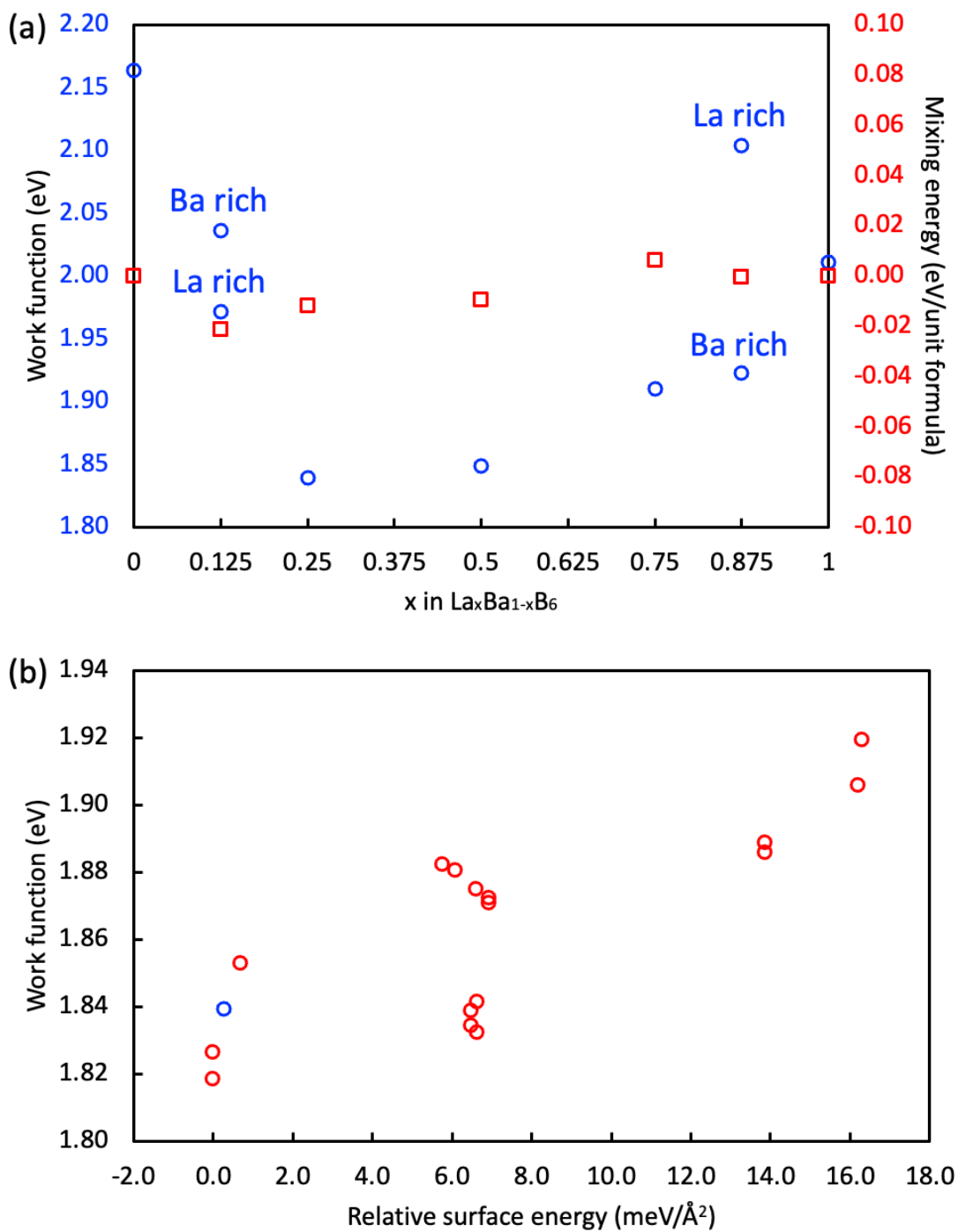

Fig 5. (a) Work functions and mixing energies of $La_xBa_{1-x}B_6$; (b) Work functions and relative surface energies (relative to lowest surface energy calculated) of $La_{0.25}Ba_{0.75}B_6$ with different surface configurations. The blue circle in panel (b) corresponds to the structure used to determine the $La_{0.25}Ba_{0.75}B_6$ data in panel (a).

Hasan *et al.* measured their lowest work function for $La_{0.31}Ba_{0.69}B_6$, very close in composition to the lowest work function we found, which was for $La_{0.25}Ba_{0.75}B_6$. Hasan *et al.* found the work function of $La_{0.31}Ba_{0.69}B_6$ to be 1.03 eV, accompanied with an extremely low Richardson constant of $A = 8.44 \times 10^{-6}$ A·cm$^{-2}$·K$^{-2}$. If the theoretical $A = 120$ A·cm$^{-2}$·K$^{-2}$ was assumed, the work function would become ~3 eV. This result makes sense, considering the emission current density at $T = 1534$ K was ~$10^{-2}$ A/cm$^2$, compared to ~0.2 A/cm$^2$ for pure $LaB_6$ at the same temperature from Ref. 1. From these results, it appears at first glance that alloying $LaB_6$ with Ba results in a higher work function. However, the hexaboride material that underwent thermionic emission test in the work of Hasan *et al.* was actually a composite mixture of polycrystalline $La_{0.31}Ba_{0.69}B_6$, $LaB_6$, and $BaB_6$. The emitter surface is thus a heterogeneous mixture of different materials which may result in domains of quite different work functions and associated patch field effects. These patch field effects are electric field interactions between low and high work function domains, and result in suppression of emission from the low work function surfaces, leading to an extracted thermionic work function that is higher than the work function of the most emissive, low work function material.[74] The very low Richardson constant for this composite material supports the fact that this emitter is highly heterogeneous, with limited surface area that was actively emitting. While these experiments do not demonstrate that $La_xBa_{1-x}B_6$ has a lower work function than $LaB_6$ or $BaB_6$, it also does not strongly suggest that it has a higher work function than these pure hexaborides, and thus may be consistent with our computational predictions. It is also possible that there is significant surface segregation in these materials, which we have not explored in the DFT studies. Overall, these experiments support the feasibility of $La_xBa_{1-x}B_6$ synthesis, and our present DFT results suggest that additional study with both DFT and experiments would be reasonable to assess if a low work function can be obtained with this alloyed material.

**3.3 Tetraboride work function trends, comparison with hexaborides**

Rare-earth tetraborides have similar crystal structure to hexaborides, where both contain rare-earth metals separated by large boron ($B_6$) octahedral cages. For tetraborides, there are also boron dimers linking the boron cages, resulting in a different boron structural framework than hexaborides, as shown in **Fig 1**. Following the same procedure as in hexaborides, we studied the composition trends of work functions of (001)-metal terminated tetraborides. Like the case of hexaborides, this metal terminated surface is expected to have the lowest work function compared to B termination as it has a positive dipole pointing out of the surface. Our DFT-calculated work functions of rare-earth tetraborides are shown in **Fig 6**, and, for ease of comparison, hexaboride work functions are also provided. Like the trend of hexaborides with composition, for tetraborides the work function tends to increase when progressing from left to right on the periodic table. This trend can again be understood in terms of metal electronegativity trends. Compared to the hexaboride trend, the changes in work function for tetraborides across the periodic table are smaller, and largely flat for Eu through Lu. The most striking difference between the hexaboride and tetraboride work functions is that the tetraboride values are higher for all compositions. From our analysis, in tetraborides we find the bonding between metal and boron to be less ionic than in hexaborides, i.e. fewer electrons are transferred from metal to boron. This lower ionicity for tetraborides makes the magnitude of the surface dipole smaller (i.e. less positive), resulting in a higher work function for tetraboride materials compared to hexaboride materials. Although Kudintseva *et al.* reported low work functions for $YB_4$ and $GdB_4$, the small reported Richardson constants and the fact that these materials showed lower current densities than their hexaboride counterparts[24] at the same temperature indicate that hexaborides are likely better electron emitters. In view of this, rare earth tetraborides are unlikely to be promising candidates for low work function electron emission cathode materials.

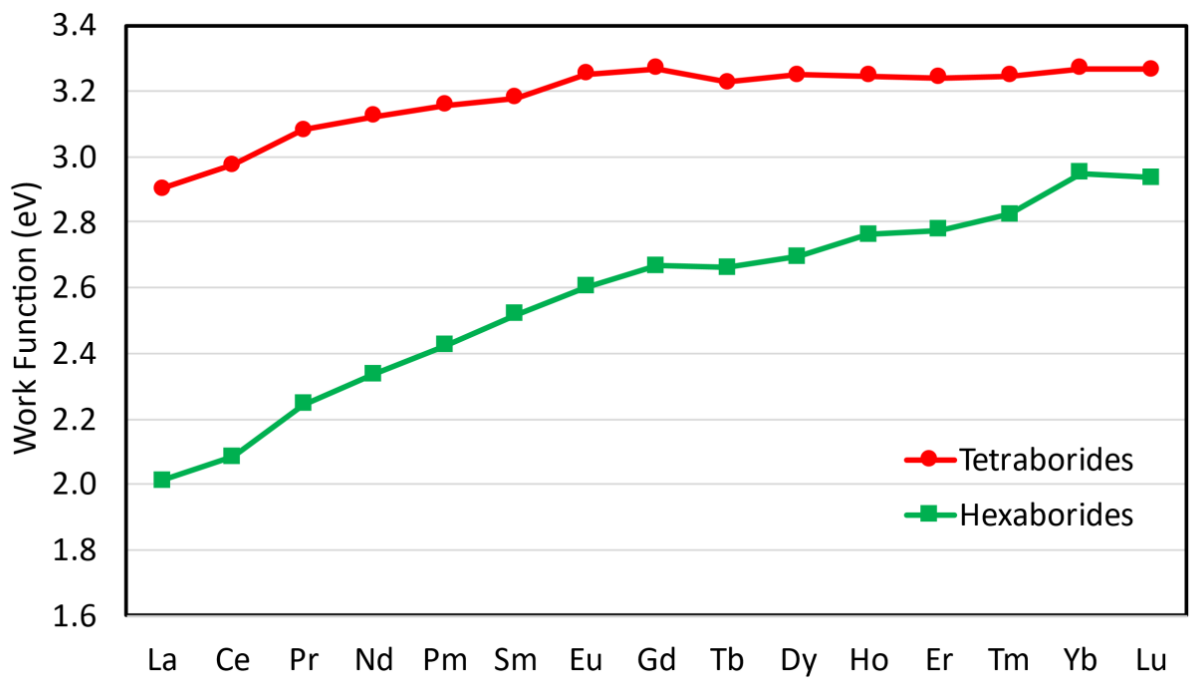

**Fig 6. The work functions of rare-earth tetraborides, for comparison, the work functions of hexaborides are also provided.**

### 3.4 Transition metal nitride work function trends, effect of surface termination

Similar to hexaborides and tetraborides, transition metal nitrides (TMNs) and transition metal carbides (TMCs) show combined properties of high melting temperature (i.e. they are refractory) and high electrical conductivity, which make them potential candidates for electron emission materials. Since experimental studies and chemical intuition suggest that TMNs possess lower work function than TMCs, here we focus on further evaluating the work function trends only for TMNs.[75]

We considered two low-index surfaces, (001) and (111), which are experimentally known to be stable at room temperature and over 500 °C.[27,28] For the (111) surface, it is possible to have surfaces terminated by either all metal atoms or nitrogen atoms, as shown in **Fig 1**. Previous computational studies suggest substantial amount of charge transfer from metal to nitrogen.[76] Considering surface dipole effects, we expect the metal terminated (111) surface to have lower work function. Calculations on select TMNs confirmed that the metal terminated (111) surface indeed has a lower work function than the nitrogen termination, therefore, we focus on metal terminated (111) surfaces and (001) surfaces for evaluating work function trends of TMNs.

In **Fig 7(a)** we show our calculated work functions of TMNs from Sc to Ni, for both (001) and (111)-metal terminations. For all TMNs in our calculations, we find that the (001) surface has lower work functions than the (111) surface. For the specific case of ZrN we have calculated the energies of the (001) and (111) (averaged over metal and nitrogen terminations) surface and find that the (001) surface has lower surface energy, which suggests that this surface may be stable in general in this class of materials. The work functions of both surfaces generally increase moving from left to right in the transition metal series, again in qualitative agreement with trends of metal electronegativity. For the (001) surface, the minimum value appears at TiN ($Ti^{3+}$ has a $3d^1$ configuration) and the maximum appears at CoN ($Co^{3+}$ has a $3d^6$ configuration). Previous computational work on $4d$ TMC work functions from Hugosson[23] showed a similar trend where NbC ($Nb^{4+}$ has a $4d^1$ configuration) had the lowest work function and PdC ($Pd^{4+}$ has a $4d^6$ configuration) had the maximum value. Since nitrogen contributes one more electron than carbon, it is thus plausible that TMN work functions instead have a minimum for TiN and maximum for CoN.

TMNs formed by early transition metals are more stable and they also have lower work functions based on our calculations. Therefore, we further calculated the work functions of ZrN, HfN and NbN, TaN to analyze the trend in work functions between the 3d, 4d, and 5d TMN materials. These nitrides are stable and have been experimentally synthesized.[25] The results are shown in **Fig 7(b)**, where TMNs of higher period have lower work functions, with HfN having the lowest work function (2.16 eV) among all the TMNs considered in our calculations. We believe the trend of lower TMN work function when progressing from 3d to 5d series can be understood as both being due to the lower electronegativity of metal elements when progressing from 3d to 5d series for these materials and the larger atomic radius of the metal atom facilitating an enhanced positive surface dipole, resulting in lower work function. With a work function of 2.16 eV, which is comparable to $LaB_6$, we propose HfN as a potential new electron emission material worthy of further study. Gotoh *et al.*[27] measured the work function of HfN thin film to be 4.7 eV by Kelvin probe, which is ~2.5 eV higher than our calculated value for (001) surface. This substantial discrepancy can be attributed to the thin film largely exposing (111) surface, which has much higher work function than (001) surface.

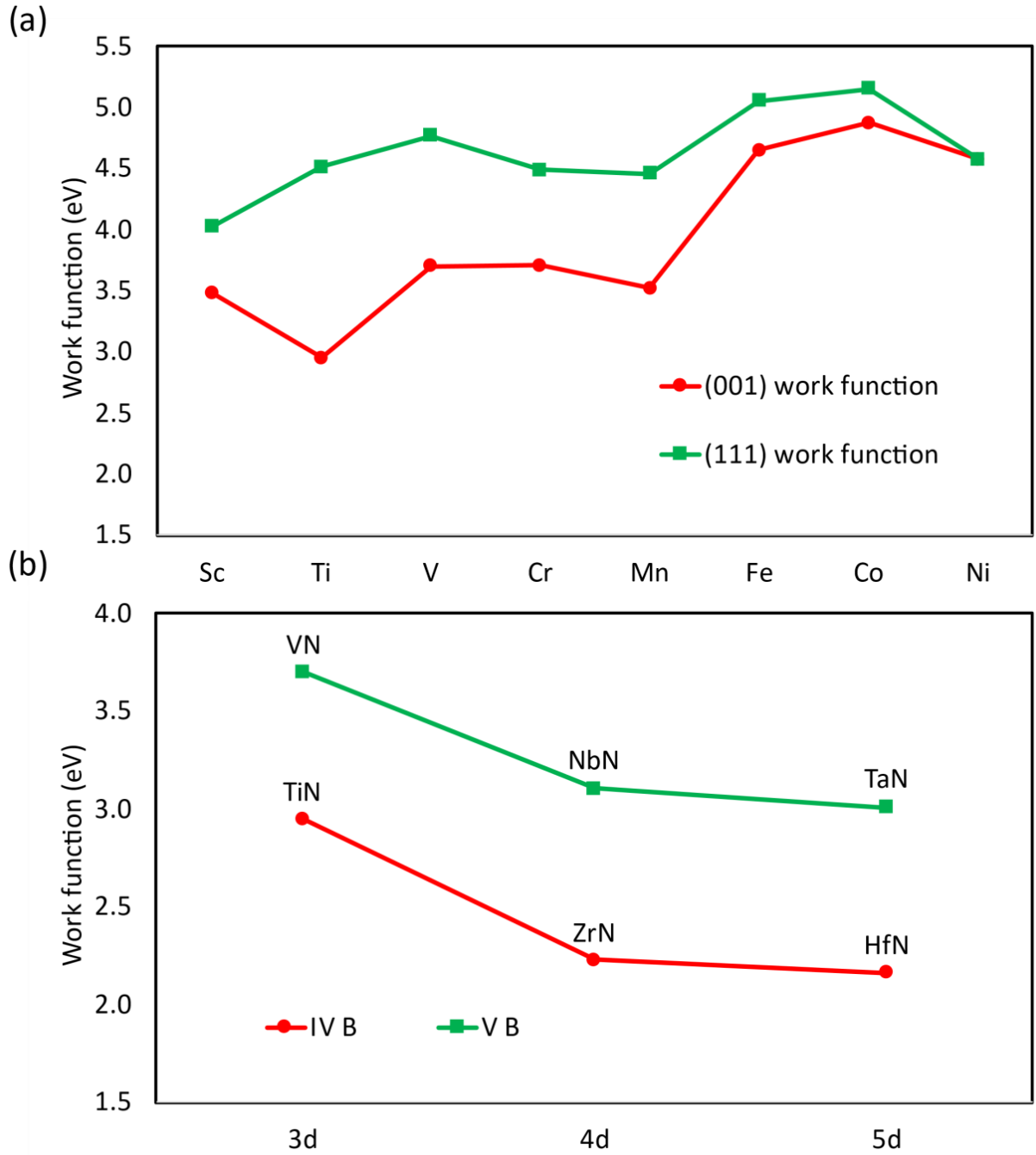

**Fig 7. The work functions of TMNs from Sc to Ni, and TMNs of IVB and VB groups.**

## 4. Summary and Conclusions

In this study, we have used Density Functional Theory (DFT) calculations to systematically study the work function trends of rare-earth and alloyed hexaborides, rare-earth tetraborides, and transition metal nitrides (TMNs). From our results, we observed increasing work functions from $LaB_6$ to $LuB_6$, and confirmed that $LaB_6$ has the lowest work function in pure rare-earth hexaborides. This trend can be understood based on the trend in electronegativity of rare-earth metals. We then explored several $La_xBa_{1-x}B_6$ alloys and found they can possess lower work functions than either $BaB_6$ or $LaB_6$. The reduced work functions of $La_xBa_{1-x}B_6$ compared to the $LaB_6$ and $BaB_6$ endmembers originate from more significant surface relaxations induced by the misfit in atomic radius of the two metal species. We thus believe $La_xBa_{1-x}B_6$ alloys are a promising electron emission material that may offer improvements compared to commercial $LaB_6$, and is worth further investigation.

Our study of rare-earth tetraboride work function trends resulted in a similar trend to hexaboride work functions. However, tetraboride work functions are all higher than their hexaboride counterparts, which we understand on the basis of differences in bonding ionicity and resulting surface dipoles between these two boride families. Due to their higher work functions, it is unlikely that tetraborides will be suitable for low work function electron emission applications.

Finally, our computational analysis of TMNs found that TMNs containing early transition metals have lower work functions (e.g. TiN) than TMNs with late transition metals (e.g. CoN). We further found that when progressing from the 3d to 5d transition metal series for early transition metals (i.e. TiN-ZrN-HfN) resulted in lower work functions progressing down the periodic table. We understand these trends as being due both to metal electronegativity trends and the influence of larger metal atomic sizes on the resulting surface dipole. These calculations resulted in the suggestion that HfN is a promising electron emission material with a low work function of 2.16 eV.

**Acknowledgements**

This work was funded by the Defense Advanced Research Projects Agency (DARPA) through the Innovative Vacuum Electronic Science and Technology (INVEST) program. This work used the Extreme Science and Engineering Discovery Environment (XSEDE),[77] which is supported by National Science Foundation grant number ACI-1548562. This work used the XSEDE Stampede2 at the Texas Advanced Computing Center (TACC) through allocation TG-DMR090023. This research was performed using the compute resources and assistance of the UW-Madison Center for High Throughput Computing (CHTC) in the Department of Computer Sciences.

**Conflicts of Interest**

There are no conflicts to declare.

**Data and Supporting Information**

The key VASP input and output files for all materials examined in this work are publicly available on Figshare at http://doi.org/10.6084/m9.figshare.14502009. In addition, a spreadsheet containing the data used to create each figure is also included.